# Testing the Early Mars $H_2$-$CO_2$ Greenhouse Hypothesis with a 1-D Photochemical Model

Natasha Batalha[a,b,c**], Shawn D. Domagal-Goldman[c,d], Ramses Ramirez[e,f,g],
James F. Kasting[b,c,h]

**Abstract.** A recent study by Ramirez et al. (2014) demonstrated that an atmosphere with 1.3-4 bar of $CO_2$ and $H_2O$, in addition to 5-20% $H_2$, could have raised the mean annual and global surface temperature of early Mars above the freezing point of water. Such warm temperatures appear necessary to generate the rainfall (or snowfall) amounts required to carve the ancient martian valleys. Here, we use our best estimates for early martian outgassing rates, along with a 1-D photochemical model, to assess the conversion efficiency of CO, $CH_4$, and $H_2S$ to $CO_2$, $SO_2$, and $H_2$. Our outgassing estimates assume that Mars was actively recycling volatiles between its crust and interior, as Earth does today. $H_2$ production from serpentinization and deposition of banded iron-formations is also considered. Under these assumptions, maintaining an $H_2$ concentration of ~1-2% by volume is achievable, but reaching 5% $H_2$ requires additional $H_2$ sources or a slowing of the hydrogen escape rate below the diffusion limit. If the early martian atmosphere was indeed $H_2$-rich, we might be able to see evidence of this in the rock record. The hypothesis proposed here is consistent with new data from the Curiosity Rover, which show evidence for a long-lived lake in Gale Crater near Mt. Sharp. It is also consistent with measured oxygen fugacities of martian meteorites, which show evidence for progressive mantle oxidation over time.

**Corresponding author**: Natasha Batalha, neb149@psu.edu

## 1. Introduction

Observations of the Martian surface reveal complex valley networks that can only be explained by running water in the distant past (Irwin et al. 2008; Grott et al. 2013). Analyses of crater morphologies (Fassett & Head 2008) suggest that this water was present circa 3.8 Ga. Further support for the warm early Mars hypothesis has been provided just recently by new data obtained by the Mars Curiosity Rover. Deposits at Gale Crater have been interpreted as being formed in a potentially habitable fluvio-lacustrine environment (Grotzinger et al., 2013), and the rover has observed stacked sediments at Mt. Sharp in Gale Crater which suggest the presence of a lake that lasted a million years or more. This implies prolonged warm conditions and a relatively Earth-like hydrologic cycle (http://mars.jpl.nasa.gov/msl/news/whatsnew/index.cfm?FuseAction=ShowNews&NewsID=1761). New estimates for the global equivalent early martian water reservoir have recently been calculated to be 137 m, and this may only be a lower limit (see Section 6.2) (Villanueva et al. 2015). That said, exactly how Mars was able to maintain an environment suitable for liquid water remains an open question, as modelers have been mostly unsuccessful at recreating these types of warm and wet conditions in their simulations.

[a]Department of Astronomy and Astrophysics, Penn State University, University Park, PA 16802, USA
[b]Center for Exoplanets and Habitable Worlds, Penn State University, University Park, PA 16802, USA
[c]NASA Astrobiology Institute Virtual Planetary Laboratory
[d]Planetary Environments Laboratory, NASA Goddard Space Flight Center, 8800 Greenbelt Road, Greenbelt, MD, 20771, USA
[e]Carl Sagan Institute, Cornell University, Ithaca, NY 14850, USA
[f]Department of Astronomy, Cornell University, Ithaca NY, 14850, USA
[g]Center for Radiophysics and Space Research, Cornell University, Ithaca NY, 14850, USA
[h]Department of Geosciences, Penn State University, University Park, PA 16802. USA

Some authors have argued that sporadic impacts during the Late Heavy Bombardment may have generated steam atmospheres and that the ensuing rainfall (about 600 m total planet wide during that entire period) carved the valley networks (Segura et al. 2002; Segura et al. 2008; Segura et al. 2012). This hypothesis seems unlikely because the amount of water required to form the valley networks is higher than that by at least three orders of magnitude, according to estimates made using terrestrial hydrologic models (Hoke et al. 2011; Ramirez et al. 2014). Extending the duration of these warm, impact-induced atmospheres is theoretically possible if cirrus clouds provide strong warming (Urata & Toon 2013); however, doing so requires high fractional cloud cover almost everywhere, and so it would be nice to see this prediction verified by independent calculations. Wordsworth et al. 2013 propose transient warming episodes caused by repeated volcanic or impact episodes, but they also find that achieving the necessary erosion rates remains challenging. Kite et al. 2013 invoke the idea that liquid water was the product of seasonal warming episodes, specifically at the equator. For seasonal melting to occur, though, there still must have been a source of precipitation and the energy to power it, so this mechanism does not resolve the issue of where the water originally came from. Most recently, Halevy & Head (2014) argued that early Mars was transiently warmed by $SO_2$ emitted during intense episodes of volcanic activity and that daytime surface temperatures at low latitudes (and low planetary obliquity) may have been high enough to result in rainfall. Their 1-D climate model may overestimate temperatures near the subsolar point, though, as it does not include horizontal heat transport. We discuss their hypothesis further in Section 6.1.2 below.

The late Noachian-early Hesperian period (~3.8 – 3.6 Ga) was also characterized by substantial weathering, as evidenced by the global distribution of phyllosilicates (e.g. clays). Although phyllosilicate formation requires long-term contact between igneous rocks and liquid water (Poulet et al. 2005; Carter et al. 2013), some investigators suggest that this process could occur in the subsurface (Ehlmann et al. 2009; Meunier et al. 2012). Hydrothermal systems could accomplish this, in principle; however, they require recharging with water, and it is unclear how this could happen if the surface was cold and dry. Other authors have argued that widespread surface clay formation is suggestive of a warmer and wetter past climate (Loizeau et al. 2010; Noe Dobrea et al. 2010; Gaudin et al. 2011; Le Deit et al. 2012; Carter et al. 2013), opposing the claim that valley network formation is the product of short climatic warming episodes (Poulet et al. 2005).

An alternative to the cold Mars hypotheses is the notion that early Mars exhibited a relatively long period of warmth characterized by a dense atmosphere dominated by greenhouse gases. Early work suggested that this could be accomplished (Pollack et al. 1987) with only $CO_2$ and $H_2O$ as greenhouse gases; however, these authors erred by neglecting condensation of $CO_2$. Subsequent climate modelers (Kasting 1991; Tian et al. 2010; Wordsworth et al. 2010; Forget et al. 2013; Wordsworth & Pierrehumbert 2013) have been unable to warm early Mars when $CO_2$ condensation is included in their simulations. However, Ramirez et al. (2014) were successful in creating above-freezing temperatures when $CO_2$-$H_2$ collision-induced absorption effects were included in their calculations. A 5% $H_2$ atmosphere with a ~3-bar 95% $CO_2$ component produced 273 K surface temperatures, and models with 10-20% $H_2$ produced temperatures above 290 K. The dense, $CO_2$-rich atmospheres required in this and other warm early Mars models have often been criticized on the grounds that they should have left extensive carbonate sediments on the surface, none of which has been observed. But the rain falling from a 3-bar $CO_2$ atmosphere would have had a pH of 3.7 or less (Kasting 2010, Ch. 8), which would almost certainly have dissolved any such rocks, allowing the carbonate to be redeposited on the subsurface. Carbonates have occasionally been detected at the bottoms of fresh craters (Michalski & Niles 2010) but most craters are likely filled with dust, and so it is not obvious that the carbonates should always show up in this type of observation.

While the Ramirez et al. work found a combination of greenhouse gases that could explain a warm and wet early Mars, the feasibility of that combination has not yet been demonstrated. A 5% $H_2$ atmosphere requires a total hydrogen outgassing rate of $8 \times 10^{11}$ $H_2$ molecules cm$^{-2}$s$^{-1}$, if hydrogen escapes at the diffusion-limited rate (see Section 2 below). Ramirez et al. made estimates of $H_2$ outgassing rates on early Mars that came within a factor of 2 of this value. This factor of 2, they argued, could be recovered if hydrogen escaped from early Mars at less than the diffusion-limited rate. However, the knowledge of the escape rate of H from the martian atmosphere is poorly constrained. Part of the problem is that we do not know how water-rich early Mars might have been. Data concerning the volatile content of the martian crust have been obtained from meteorite (Kurokawa et al. 2014) and in situ (Mahaffy et al., 2015) analyses, but they still leave an order of magnitude uncertainty in the global near-surface water inventory prior to 4 Ga. While a higher escape rate could be offset by a higher volcanic $H_2$ outgassing rate or by supplementing volcanic $H_2$ with other $H_2$ sources, this idea has not been previously explored.

In this paper, we test the plausibility of the high-$H_2$ hypothesis of Ramirez et al. (2014) by using a photochemical code to study whether such an atmosphere would be sustainable over geological timescales. We do this by carefully maintaining the redox balance of each simulation, looking for self-consistent atmospheres that could maintain liquid water at the planet's surface. These simulations allow us to infer the volcanic fluxes required to maintain the high $H_2$ levels needed to keep early Mars warm. We also consider the potential climatic effects of species other than $CO_2$, $H_2O$, and $H_2$ – specifically CO, $CH_4$, $SO_2$, and $H_2S$. Finally, we consider whether the $H_2$ greenhouse hypothesis might be tested using Mars rover, orbiter, and meteorite data.

## 2. The atmospheric and global redox budgets

All atmospheres must be in approximate redox balance over sufficiently long time scales; otherwise, their oxidation state would change during the time frame of interest. For an $H_2$-rich atmosphere, 'long' means time scales of tens to hundreds of thousands of years (Kasting, 2013). Both an atmospheric redox budget and a global redox budget can be computed (see, e.g., Kasting & Canfield 2012; Kasting, 2013). The global redox budget is defined as the redox budget of the combined atmosphere-ocean system. This is the budget that is considered in models of the modern Earth's redox balance (e.g., Holland, 2002, 2009).

To balance the atmospheric redox budget, we assume that the sources of reducing power to the atmosphere are volcanic outgassing, $\phi_{out}(Red)$, and rainout/surface deposition of oxidizing species, $\phi_{rain}(Oxi)$. The sources of oxidizing power are rainout of reduced species, $\phi_{rain}(Red)$, and the escape of hydrogen to space, $\phi_{esc}(H_2)$. Given these definitions, a balanced atmospheric redox budget should obey the following relationship:

$$\phi_{out}(Red) + \phi_{rain}(Oxi) = \phi_{esc}(H_2) + \phi_{rain}(Red) \qquad [1]$$

Typically, our atmospheric photochemical model balances the redox budget to about 1 part in $10^7$. The escape rate of hydrogen is given by the diffusion-limited expression (Walker, 1977)

$$\Phi_{esc}(H_2) = \frac{b_i}{H_a} \frac{f_T(H_2)}{1 + f_T(H_2)} \cong \frac{b_i f_T(H_2)}{H_a} \qquad [2]$$

Here, $b_i$ is the weighted molecular diffusion coefficient, $H_a (= kT/mg)$, is the scale height (at T~160K, $\frac{b_i}{H_a} = 1.6 \times 10^{13} \mathrm{cm^{-2} s^{-1}}$ ) and $f_T(H_2)$ is the total hydrogen volume mixing ratio: $f_T(H_2) = f(H_2) + 0.5f(H) + f(H_2O) + 2f(CH_4) + \ldots$, expressed in units of $H_2$ molecules.

Realistic models must also balance the redox budget of the combined atmosphere-ocean system. Kasting (2013) refers to this as the *global redox budget*. His expression for this budget is as follows:

$$\Phi_{out}(Red) + \Phi_{OW} + \Phi_{burial}(CaSO_4) + \Phi_{burial}(Fe_3O_4)$$
$$= \Phi_{esc}(H_2) + 2\Phi_{burial}(CH_2O) + 5\Phi_{burial}(FeS_2)$$
$$[3]$$

Here, $\Phi_{OW}$ represents oxidative weathering of the continents and seafloor, and $\Phi_{burial}(i)$ is the burial rate of species $i$. $H_2O$, $CO_2$, $N_2$, and $SO_2$ are taken as redox neutral species in this formulation.

The global redox balance takes into account processes occurring at the ocean-sediment interface, *e.g.*, burial of organic carbon and pyrite. If we assume, as a starting point, that nothing is happening at that interface, and if oxidative weathering is neglected, then the global redox budget simplifies to

$$\phi_{out}(Red) = \phi_{esc}(H_2)$$

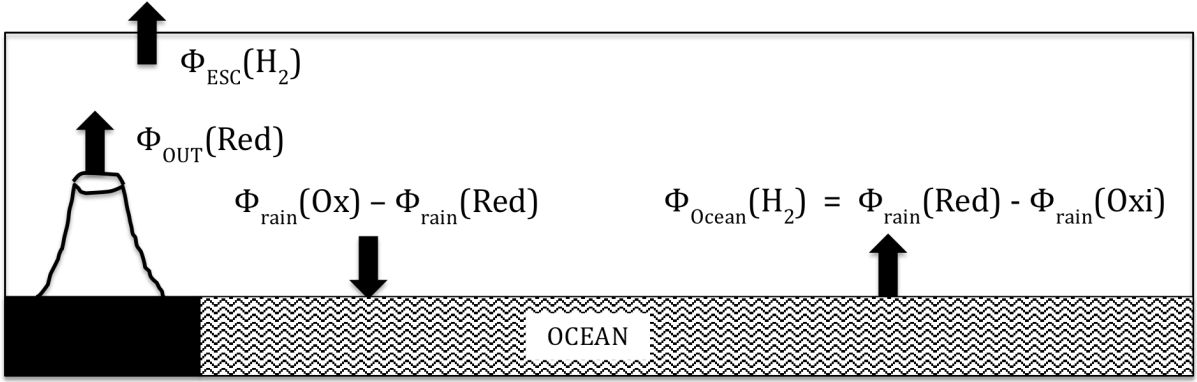

**Figure 1**. A schematic diagram showing the method used for balancing the ocean-atmosphere system. $\Phi_{out}$(Red) is the flux of reductants outgassed through volcanoes, $\Phi_{rain}$(Red/Ox) is the flux of rained-out reductants or oxidants (including surface deposition). $\Phi_{Ocean}$ ($H_2$) is the flux of $H_2$ back into the atmosphere required to balance the oceanic $H_2$ budget. An excess of reductants, such as $H_2S$, flowing into the ocean leads to an assumed upward flux of $H_2$.

[4]

*i.e.,* volcanic outgassing of reduced gases must be balanced by escape of hydrogen to space. Inserting this equation back into Eq. [1] implies that

$$\phi_{rain}(Oxi) = \phi_{rain}(Red).$$

[5]

that is, the rainout rate of oxidants from the photochemical model must equal the rainout rate of reductants. Or, to think of this in a different way, if no redox reactions are happening on the seafloor, the rate of transfer of oxidants from the atmosphere to the ocean must equal the rate of transfer of reductants. We will start from this simplifying assumption and then add seafloor processes as we proceed.

In practice, a photochemical model will *not* satisfy Eqs. [4-5] on its own. Rather, the photochemical modeler must make decisions about how to deal with any imbalance. Although previous models have typically not considered the ocean-atmosphere balance (Segura et al. 2003; Tian et al. 2010), two more recent models have done so (Domagal-Goldman et al. 2014; Tian et al. 2014). Domagal-Goldman et al. (2014) did this by wrapping the photochemical code in a separate script that repeatedly ran the model, changing the boundary conditions between simulations, until the model satisfied the above equations for a specified value of $\phi_{rain}(Oxi)$ - $\phi_{rain}(Red)$ to within a determined tolerance level.

We used the simpler procedure previously employed by Tian et al. (2014). In all of our simulations, we found that $\phi_{rain}(Oxi)$ - $\phi_{rain}(Red)$ was < 0, that is, the rainout rate of reductants exceeded that of oxidants. So, we let $H_2$ flow back from the ocean into the atmosphere at a rate equal to that of the difference between the rained out reductants and oxidants. Without this assumption, $H_2$ would flow back into the planet (without any physical justification), and we might therefore underestimate the atmospheric $H_2$ concentration. The global redox budget is illustrated in Figure 1.

By following this methodology, we essentially assume that dissolved reductants and oxidants react with each other in solution in such a way as to yield $H_2$, and that the organic carbon burial rate is essentially zero. This may not always be the case, and we must be conscious of that in our analysis. For example, if the burial rate of organic carbon or other reduced species exceeded the sum of the burial rate of oxidized species and the rate of oxidative weathering, our assumptions would not apply, and atmospheric $H_2$ concentrations would decrease.

If all of the $H_2$ in the atmosphere came directly from volcanic outgassing, Eq. [2] and [4] taken together show that in order to have $f_T(H_2) = 0.05$, the minimum $H_2$ concentration needed to sustain a warm early Mars, the outgassing rate of $H_2$ must be at least $8 \times 10^{11}$ cm$^{-2}$s$^{-1}$. If an appreciable fraction of the atmospheric hydrogen is in some other form, *e.g.*, $CH_4$, then the total hydrogen outgassing rate would need to be correspondingly higher because the main greenhouse warming is coming from just the $H_2$.

This approach allows us to consider the various sources of $H_2$, including both outgassing terms and terms at the ocean-sediment interface. We outline these sources below. First, we consider contributions to $\phi_{out}(Red)$ from outgassed $H_2$, S, and $CH_4$. Then, we consider the contributions to $\phi_{rain}(Oxi)$ - $\phi_{rain}(Red)$ (assumed > 0) from serpentinization and burial of iron oxides.

## 3. Possible sources of hydrogen on early Mars

### 3.1) Volcanic Outgassing

The term 'outgassing' refers to release of gases from magma. Volcanic outgassing rates on early Mars have frequently been estimated by looking at surface igneous rocks, evaluating their ages, and making assumptions about the volatile content of the lava from which they formed (e.g., Greeley and Schneid, 1991: Grott et al., 2011; Craddock and Greeley, 2009 and refs. therein). Grott et al. (2011) estimated that 0.25 bars of $CO_2$ and 5-15 m of $H_2O$ were outgassed on Mars during the interval 3.7-4.1 Ga. The corresponding implied outgassing rates are 0.06 Tmol/yr for $CO_2$ and 0.3 Tmol/yr for $H_2O$. By comparison, the estimated outgassing rates for $CO_2$ and $H_2O$ on modern Earth are 7.5 Tmol/yr and 102 Tmol/yr, respectively (Jarrard, 2003). Even taking into account the 4 times larger surface area of Earth, the implied per unit area martian outgassing rates are 25-100 times smaller. Such outgassing rates are almost certainly too small to maintain a warm, dense atmosphere, leading some to conclude that the martian atmosphere has always been thin and cold (e.g. Forget et al. 2013; Wordsworth et al., 2013; Grott et al., 2013), except, perhaps in the aftermath of repeated explosive eruptions (Wordsworth et al., 2013) or giant impacts (Segura et al., 2002).

However, these outgassing estimates for early Mars are potentially underestimated, because they ignore the effects of volatile recycling. For example, Earth's relatively high outgassing rates result from volatile recycling between the crust and the mantle, not from juvenile degassing. Much higher outgassing rates could be expected on early Mars if the planet experienced plate tectonics and associated element recycling. Heat flow on early Mars is thought to have been comparable to that on modern Earth (Montesi & Zuber 2003), so some authors have postulated that outgassing rates of major volatiles may also have been similar (Ramirez et al., 2014, Halevy & Head, 2014). Evidence for past tectonic activity incudes Mars Global Surveyor data of an alternating polarity in the remanent magnetic field, inferred to be evidence for sea-floor spreading (Connerney 1999), and major faults associated with these crustal variations. Magnetic polarity patterns deduced by Connerney (1999) are Noachian in age, but Sleep (1994) suggests that plate tectonics extended at least through the early Hesperian. Along these same lines, Anguita et al. (2001) discuss how plate tectonics better explains the observed tectonic regime in the early Hesperian than do other hypotheses. No consensus has been reached on this topic, however; for example, Grott et al. (2013) have argued that crust-mantle recycling never occurred on Mars because plate tectonics never got started. Although the notion that plate tectonics may have operated on early Mars remains controversial, it is required to support the thick, $H_2$-rich atmospheres proposed by Ramirez et al. Therefore, we assume here that Mars *did* recycle volatiles and that the overall efficiency of recycling was comparable to that on modern Earth. Future exploration of Mars will reveal whether this assumption is correct. We also consider other sources of hydrogen from oxidation of crustal ferrous iron and from photochemical oxidation of other reduced gases.

### 3.1.1 $H_2$

Differences in the composition of volatiles outgassed on Mars compared to Earth should result from the different oxidation states of their respective mantles. Earth's upper mantle is thought to have an average oxygen fugacity, $fO_2$, near that of the QFM (quartz-fayalite-magnetite) synthetic buffer. At typical surface outgassing conditions (1,450 K, 5 bar pressure), this yields $fO_2 \cong 10^{-8.5}$ (Frost et al. 1991; Ramirez et al. 2014). Given this value for $fO_2$, the $H_2$:$H_2O$ ratio, $R$, in the gas that is released can be calculated from the expression

$$\frac{P_{H_2}}{P_{H_2O}} \equiv R = \left(\frac{K_1}{fO_2}\right)^{0.5}, \qquad [6]$$

Here, $P_{H_2}$ and $P_{H_2O}$ are the partial pressures of the two gases, and $K_1$ (=1.80× $10^{-12}$ atm) is the equilibrium constant for the reaction: $2H_2O \overset{K_1}{\Longleftrightarrow} 2H_2 + O_2$ (Ramirez et al., 2014). Plugging $K_1$ and the terrestrial mantle $f_{O_2}$ into Eq. [6] gives an $H_2:H_2O$ ratio of 0.024. This yields a terrestrial $H_2$ outgassing flux of ~2.4 Tmol/yr, when multiplied by the terrestrial $H_2O$ subaerial outgassing rate of 100 Tmol/yr, or ~$3.7 \times 10^{11}$ cm$^{-2}$s$^{-1}$ (Jarrard 2003). (On Earth, the conversion from geochemical to atmospheric science units is 1 Tmol/yr = $3.74 \times 10^9$ cm$^{-2}$s$^{-1}$.) The terrestrial $H_2$ outgassing rate is therefore of the order of $1 \times 10^{10}$ cm$^{-2}$s$^{-1}$, with a factor of 2 or more uncertainty in either direction (Holland 2009).

Mars' oxygen fugacity is thought to be at least 3 log units lower than Earth's, near ~ IW+1 (Grott et al. 2011). IW is the iron-wüstite buffer, which has an $fO_2$ about 4 log units below QFM. Based on this observation, Ramirez et al. (2014) calculated that Mars could have outgassed $H_2$ at up to 40 times the rate of Earth: $40 \times 10^{10}$ cm$^{-2}$s$^{-1}$ = $4 \times 10^{11}$ cm$^{-2}$s$^{-1}$. This estimate included a 50 percent contribution from $H_2S$, which was assumed to be oxidized to $SO_2$ by atmospheric photochemistry, according to

$$H_2S + 2H_2O \rightarrow SO_2 + 3H_2 \qquad [7]$$

We demonstrate in the next section that this assumption may be unfounded, and so outgassing of $H_2S$ may not have added much to Mars' atmosphere. That said, terrestrial $H_2$ outgassing estimates are uncertain by about a factor of 2 or more, and the early Martian mantle could have had an oxygen fugacity near IW-1 (Warren & Gregory 1996). The latter factor alone could have approximately doubled the estimated $H_2$ outgassing rate (Ramirez et al., 2014). Thus, an $H_2$ outgassing rate of $8 \times 10^{11}$ cm$^{-2}$s$^{-1}$ is not implausible; it just requires slightly more optimistic assumptions than have hitherto been adopted. Specifically, this higher outgassing rate is highly dependent on the estimate of the redox state of the ancient Martian mantle, and is the single largest source of uncertainty in our estimates of Martian $H_2$ outgassing rates. This highlights the importance of future measurements that might reduce the uncertainties in that quantity.

### 3.1.2 Sulfur

Several hundred millibar to as much as 1 bar of sulfur may have been outgassed via juvenile degassing throughout Martian history (Craddock & Greeley 2009). This leads to a sulfur outgassing rate of at most $6 \times 10^6$ cm$^{-2}$s$^{-1}$, if we assume it was outgassed over a period of ~1 billion years. As with other estimates of juvenile degassing, this small outgassing rate is not enough to maintain a warm, dense atmosphere on early Mars. On Earth, volcanic sulfur comes from three main sources: arc volcanism, hotspot volcanism, and submarine volcanism.

Direct satellite measurements of arc volcanism yield $SO_2$ outgassing rates of 0.2-0.3 Tmol/yr (equivalent to ~$(0.7-1.1) \times 10^9$ cm$^{-2}$s$^{-1}$ via the conversion above) (Halmer et al. 2002). These numbers probably underestimate the total $SO_2$ outgassing rate, as they only measure the $SO_2$ outgassed through explosive volcanism. Instead, if we combine the ratio of total sulfur to $H_2O$ in arc volcanism (~0.01 in Fig 6, Holland 2002) and the ratio of $H_2O$ to $CO_2$ (~30 in Fig 6, Holland 2002), then the rate of $SO_2$ outgassing on modern Earth should be ~0.8 Tmol/yr, assuming a $CO_2$ outgassing rate from arc volcanism of ~2.5 Tmol/yr (Jarrard 2003).

Hotspot volcanism, such as that which occurs in Hawaii, also contributes to sulfur outgassing. Although hotspot outgassing rates are difficult to accurately determine, the observed ratio of $SO_2/CO_2$ is ~0.5 (Walker 1977, Table 5.5). Therefore, if the release rate of carbon from hotspot volcanism on Earth is 2 Tmol/yr (Jarrard 2003), the corresponding release rate of $SO_2$ should be ~1 Tmol/yr. This leads to a total subaerial $SO_2$ outgassing rate of 1.8 Tmol/yr or ~$6.7 \times 10^9$ cm$^{-2}$s$^{-1}$.

Sulfur is also outgassed as $H_2S$ during submarine volcanism. Holland (2002) averaged measurements of $H_2S$ concentrations in hot, axial vent fluids, 3-80 mmol/kg (Von Damm 1995; Von Damm 2000), to estimate dissolved $H_2S$ concentrations of 7 mmol/kg. By combining this value with estimates for the total emergent water flux, ~$5 \times 10^{13}$ kg/yr, we can convert the $H_2S$ concentration to an $H_2S$ outgassing rate of 0.35 Tmol/yr, or ~$1.3 \times 10^9$ cm$^{-2}$s$^{-1}$.

Gaillard & Scaillet (2009) show that on Mars, $H_2S$ and $SO_2$ should be outgassed at approximately the same rate for a mantle redox state near IW. Therefore, we use an outgassing rate of $5 \times 10^9$ cm$^{-2}$s$^{-1}$ for both $SO_2$ and $H_2S$, which is roughly in agreement with the values above. We note parenthetically that Halevy and Head (2014) assumed that all sulfur outgassed on early Mars was in the form of $SO_2$.

### 3.1.3 Carbon

On modern Earth carbon is outgassed as $CO_2$ at a rate of ~7.5 Tmol/yr ($2.8 \times 10^{10}$ cm$^2$s$^{-1}$) (Jarrard 2003). We expect that carbon outgassing should have also been a major contributor to the early martian atmosphere. A recent study by Wetzel et al. (2013) showed that carbon should be stored in different forms in planetary mantles, depending on the oxygen fugacity, $fO_2$. At $fO_2$ values above IW-0.55, carbon is stored as carbonate in the melt and would be outgassed as $CO_2$. At $fO_2$ values below IW-0.55, carbon is stored as iron carbonyl, $Fe(CO)_5$, and as $CH_4$ and would be outgassed as CO and $CH_4$. Wetzel et al. (2013) calculated that initial solidification of a 50 km-thick crust should lead to outgassing of 1.3 bar $CH_4$ and 1 bar CO at low $fO_2$ or to 11.7 bar of $CO_2$ at higher $fO_2$. The higher pressure of the outgassed $CO_2$ atmosphere is caused by a factor of two increase in carbon solubility in melts at $fO_2 >$ IW-0.55, along with the higher molecular weight of $CO_2$ compared to CO and $CH_4$. Assuming this gets released over ~1 billion years, we derive a lower bound estimate for carbon outgassing of $6 \times 10^6$ cm$^2$s$^{-1}$ at low $fO_2$ or about twice that value at higher $fO_2$. As pointed out earlier, juvenile outgassing rates are always relatively small.

A low rate of carbon outgassing may not be an issue for the $CO_2$ content of the early martian atmosphere because the $CO_2$ removal rate from silicate weathering depends on temperature. At low surface temperatures, liquid water would not be present and so $CO_2$ would not be lost by this process. 11.7 bar of $CO_2$, or even half that amount, is more than adequate for the greenhouse atmospheres postulated here. $CO_2$ could also have been lost by solar wind interactions, as happens today, but such loss might have been precluded if Mars had a magnetic field at this time. We assume that the early martian atmosphere was not being rapidly stripped away in this manner.

Here, we are interested in whether outgassing of carbon in reduced form could have provided an additional source of $H_2$. Outgassing rates *do* matter in this case because hydrogen is always being lost to space, regardless of the presence or absence of a magnetic field. At low mantle $fO_2$ values, outgassing of the reduced gases CO and $CH_4$ could have contributed to the atmospheric $H_2$ budget. The outgassing rate computed from initial crustal solidification would not have been high enough to supply an appreciable amount of additional $H_2$. To obtain a higher estimate, consistent with our estimate for direct $H_2$ outgassing above, we used modern Earth carbon outgassing estimates (Jarrard 2003) and assumed that mantle $fO_2$ was $<$ IW-0.55 and that CO and $CH_4$ were released in the 1:1.3 ratio calculated by Wetzel et al. This yields outgassing rates of $2 \times 10^{10}$ cm$^{-2}$s$^{-1}$ and $8 \times 10^9$ cm$^{-2}$s$^{-1}$ for $CH_4$ and CO, respectively. If all of the $CH_4$ was oxidized to $CO_2$ following the stoichiometry

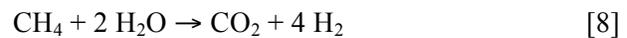

$$CH_4 + 2\ H_2O \rightarrow CO_2 + 4\ H_2 \qquad [8]$$

the equivalent $H_2$ production rate should have been 4 times the $CH_4$ outgassing rate, or $8 \times 10^{10}$ cm$^{-2}$s$^{-1}$. This is about 1/10[th] the $H_2$ flux needed to sustain a warm $H_2$-$CO_2$ greenhouse atmosphere. CO outgassing is less important as a source of $H_2$, as its outgassing rate is lower and its stoichiometric coefficient for $H_2$ production, assuming oxidation to $CO_2$, is only unity

$$CO + H_2O \rightarrow CO_2 + H_2. \qquad [9]$$

### 3.2 Serpentinization

A second possible source of hydrogen to Mars' early atmosphere is serpentinization. This process differs from volcanic outgassing in that it occurs at relatively low temperatures, 500-600 K, whereas outgassing from magmas occurs at the melt temperature of ~1450 K. Serpentinization occurs when warm water interacts with ultramafic (Mg- and Fe-rich) basalts. Ferrous iron contained in the basalts is excluded from the serpentine alteration products, and so it forms magnetite, releasing $H_2$ in the process

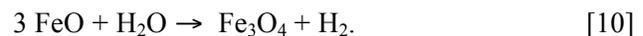

$$3\ FeO + H_2O \rightarrow\ Fe_3O_4 + H_2. \qquad [10]$$

Evidence for serpentinization on Mars exists in the form of ultramafic rocks discovered on the Martian surface. Olivine concentrations of $10 - 20\%$ have been detected both by the Thermal Emission Spectrometer (Koeppen & Hamilton 2008) and in SNCs (e.g. McSween et al. 2006) . Moreover, the Mars Reconnaissance Orbiter (MRO) has detected serpentine itself from orbit (Ehlmann et al. 2009).

Serpentinization is a minor source of hydrogen to Earth's current atmosphere, accounting for ~0.4

Tmol $H_2$/yr, or $1.5 \times 10^9$ cm$^{-2}$s$^{-1}$ (Sleep 2005; Kasting, 2013). For this process to have made an important contribution to the early martian $H_2$ budget, it would have needed to occur 10-100 times faster than it does on Earth today. That sounds daunting, but it may be possible. Most oceanic basalts today are *not* prone to serpentinization because the terrestrial seafloor is predominantly mafic, not ultramafic. Ultramafic rocks, e.g., peridotites, are found deep within the seafloor and are exposed to hydrothermal circulation within slow-spreading ridges such as the Mid-Atlantic Ridge. Earth's upper mantle should have been hotter in the past; hence, the degree of partial melting during seafloor creation should have been higher, and the seafloor itself should have been more mafic, or even ultramafic. Similarly, if Mars' upper mantle was originally hot, and if seafloor was being generated there as it is here on Earth, interaction of ultramafic rocks with water may have been commonplace, as well.

One can make a crude estimate of the $H_2$ flux that might have been generated by this process by drawing an analogy to seafloor oxidation on Earth today. The rate at which ferric iron is generated and carried away by seafloor spreading today is about $21 \times 10^3$ kg/s, or $1.2 \times 10^{13}$ mol/yr (Lécuyer & Ricard 1999). Most of this ferric iron is produced by sulfate reduction, not by serpentinization. But if the oceanic crust were more ultramafic, and if this same amount of ferric iron were generated by serpentinization, then according to reaction [10] it would produce 1 mole of $H_2$ for every 2 moles of ferric iron (because $Fe_3O_4$ contains two atoms of ferric iron), and so the corresponding $H_2$ flux would be 6 Tmol/yr, or $2.2 \times 10^{10}$ cm$^{-2}$s$^{-1}$. That is roughly 10% of the volcanic $H_2$ outgassing rate estimated by Ramirez et al. (2014) for early Mars with a mantle $fO_2$ near IW+1. So, unless martian seafloor was serpentinizing much faster than terrestrial seafloor gets oxidized today, this process would have been a relatively minor term in the martian $H_2$ budget. When we do the estimate this way, serpentinization appears to be a relatively minor source of $H_2$. Other authors however, have made more generous estimates of $H_2$ production from this process, as high as 35 Tmol/yr (on Mars), or $4 \times 10^{11}$ cm$^{-2}$s$^{-1}$ (Chassefière et al. 2014), about half the flux needed to sustain a 5% $H_2$ mixing ratio. So we should not rule out serpentinization as an important source of hydrogen on early Mars.

### 3.3 Photochemical Fe oxidation

On Earth, Fe oxidation by way of UV irradiation of surface waters could have also been a source of $H_2$ and could have contributed to the deposition of banded iron-formations (BIFs) (Braterman et al. 1983). Additionally, Hurowitz et al. (2010) showed that the sedimentary rocks found at Meridiani Planum on Mars were formed in the presence of acidic surface waters and that Fe oxidation may have played a role in maintaining that high acidity. This mechanism could potentially have produced large amounts of gaseous $H_2$. Still, it is uncertain how much of the martian surface was producing $H_2$ in this manner, as Meridiani Planum has an area of $\sim 2 \times 10^5$ km$^2$, only $\sim 0.1\%$ the total surface area of Mars (Hurowitz et al. 2010).

To get around this problem, we once again make an analogy to early Earth. Kasting (2013) estimated $H_2$ production rates from deposition of BIFs on the Archean Earth. His estimates ranged from (0.2-25) Tmol($H_2$)/yr, or $(0.7-9) \times 10^{10}$ cm$^{-2}$s$^{-1}$. But the higher end of this range is an extremely generous upper bound which would require dissolved $Fe^{+2}$ concentrations in vent fluids that were hundreds of times higher than those in modern terrestrial hydrothermal systems. Even with those assumptions, this mechanism would likely have been only a minor source of $H_2$ on early Mars. Despite its apparent lack of importance, we parameterize these potential effects below, because of their possible relationship to sedimentary layers on Mars, including the hematite beds on Mount Sharp in Gale Crater.

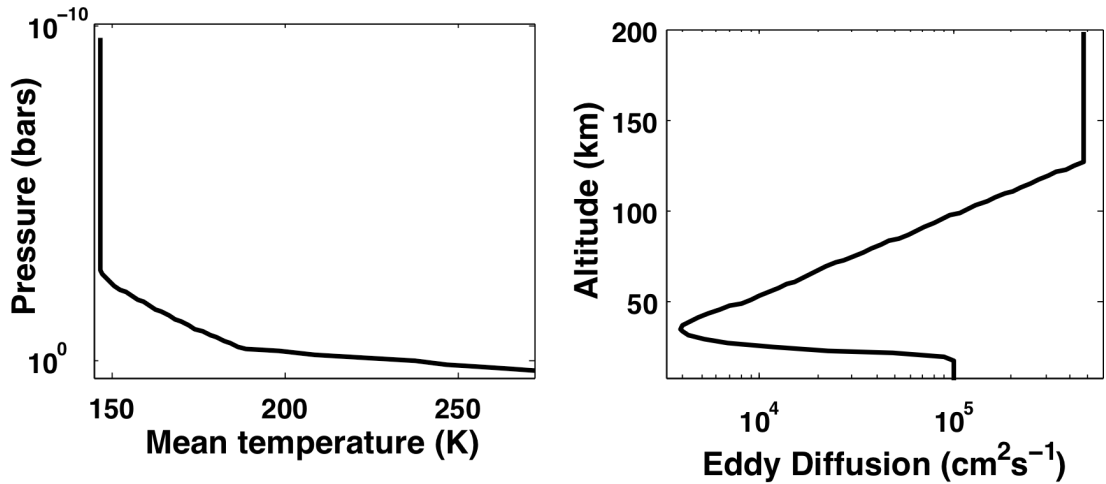

**Figure 2.** Temperature-pressure profile (top) and eddy diffusion profile (bottom) assumed for the photochemical calculations. The temperature decreases from 273K to 147K at an altitude of 67 km and then is isothermal to the top (200 km altitude). This is consistent with the 5% $H_2$, 95% $CO_2$ 3-bar atmosphere from Ramirez et al. 2014.

## 4. Model Description

### 4.1 Photochemical Model

To investigate how fast volcanic gases such as $H_2S$ and $CH_4$ would be converted into $H_2$, we used a 1-D (in altitude), horizontally averaged photochemical model that solves the coupled continuity and flux equations for multiple atmospheric species using an implicit, reverse Euler integration technique. The model, originally developed by (Kasting et al. 1979), has been most recently updated by (Domagal-Goldman et al. 2011). The model used for this study does not include higher hydrocarbon chemistry (alkenes, alkynes, alkanes longer than C2), as it is not important for the current calculation. Instead, the model includes 29 long-lived species and 16 short-lived species involved in 215 reactions (see Appendix A).

The model consists of 100 plane parallel layers spaced by 2 km in altitude, allowing it to calculate species profiles up to 200 km. We begin by assuming that Mars was wet and warm with a surface temperature of 273 K. These parameters were chosen to be consistent with a temperature-pressure profile derived by Ramirez et al. 2014 for a 3-bar, 5% $H_2$, 95% $CO_2$ atmosphere. This assumed composition ignores the possible presence of higher amounts of $N_2$ in the early martian atmosphere, which can be inferred from the high measured $^{15}N/^{14}N$ ratio today (Fox 1993). Higher $N_2$ should not greatly affect the climate; indeed, $N_2$ can substitute for $CO_2$ to create pressure-induced absorption by $H_2$ (Ramirez et al., 2014, Fig. 2). Whether our assumed composition is an accurate representation of the early martian atmosphere depends, of course, on the validity of the Ramirez et al. hypothesis. However, the point of this study is to see if such an atmosphere is sustainable, so the use of it here is consistent with that goal. This means atmospheres that do not reproduce the Ramirez et al. $H_2$ concentrations have an inconsistent temperature profile; however any simulations that could maintain such an atmosphere would be self-consistent. The temperature is assumed to decrease from 273 K at the surface to 147 K at an altitude of 67 km, following a moist $H_2O$ adiabat in the lower troposphere (0-20 km) and a moist $CO_2$ adiabat above that (20-120 km). Above 67 km, the atmosphere is assumed to be isothermal up to 200 km altitude, consistent with the assumed lack of oxygen and ozone. Several reactions rates are positively correlated with temperature (i.e. an increase in temperature causes an increase in the rate of a reaction). In the case of water vapor, a 10 K increase in temperature doubles the water vapor volume-mixing ratio. $H_2$, however, is less affected (1% increase) by this same change in temperature. Figure 2 shows this temperature profile along with the eddy diffusion profile.

### 3.2 Boundary Conditions

At the top of the atmosphere, $CH_4$, H and $H_2$ are assumed to diffuse upward at the diffusion-limited velocity (Walker 1977), while CO and O are given constant downward fluxes that balance photolysis of $CO_2$ above the top layer of our model. All other species are assigned a flux of zero at the top of the atmosphere, implying that nothing else is escaping besides hydrogen. This assumption is consistent with the presence of a magnetic field to prevent solar wind stripping and with hydrodynamic escape rates for heavy species that were slower than those calculated by (Tian et al. 2009)(whose escape model did not include appreciable concentrations of $H_2$).

*Table 1. Deposition Velocities at Lower Boundary*

| Species | Deposition Velocity (cm/s) |
|---------|---------------------------|
| O | 1 |
| $O_2$ | 0 |
| $H_2O$ | 0 |
| H | 1 |
| OH | 1 |
| $HO_2$ | 1 |
| $H_2O_2$ | 0.2 |
| CO | $1 \times 10^{-8}$ |
| HCO | 1 |
| $H_2CO$ | 0.1 |
| $CH_4$ | 0 |
| $CH_3$ | 1 |
| $C_2H_6$ | $1 \times 10^{-5}$ |
| NO | $3 \times 10^{-4}$ |
| $NO_2$ | $3 \times 10^{-3}$ |
| HNO | 1 |
| $H_2S$ | 0.015 |
| HS | $3 \times 10^{-3}$ |
| S | 1 |
| HSO | 1 |
| SO | $3 \times 10^{-4}$ |
| $SO_2$ | 1 |
| $NH_3$ | 0 |
| $NH_2$ | 0 |
| N | 0 |
| $N_2H_4$ | 0 |
| $N_2H_3$ | 0 |
| $H_2SO_4$ | 0.2 |

At the lower boundary, every species except for $H_2$ was given a constant deposition velocity. (As stated in Section 2, $H_2$ was assigned a constant upward flux at the lower boundary to ensure redox balance.) This accounts for their reaction with surface rocks and any ocean that might have been present. Table 1 lists the assumed deposition velocities for each species. Our results are insensitive to most of these deposition velocities, with the notable exception of CO. In most of our simulations, the CO deposition velocity is fixed at $10^{-8}$ cm $s^{-1}$, the value derived for an abiotic early Earth (Kharecha et al. 2005). The assumption here is that dissolved CO equilibrates with formate, but that a small percentage of that formate is photochemically converted to acetate and is lost from the atmosphere-ocean system. This implies that there is a small, but finite, burial flux of organic carbon (as acetate). Sensitivity tests, described below, were performed to determine the effect of varying the CO deposition velocity.

### 4.3 Volcanic Outgassing

Six different gases, $H_2$, CO, $CH_4$, $NH_3$, $SO_2$, and $H_2S$, were assumed to have sources from volcanic outgassing. $CO_2$ has a volcanic source, as well, but the $CO_2$ mixing ratio is fixed in our model at 0.95. The other 5% of the atmosphere is assumed to consist of $N_2$. When $H_2$ builds up to appreciable concentrations in the model, it displaces $CO_2$. Volcanic fluxes for the base case and final model are listed in Table 2. These fluxes were distributed over the lowest 20 km of the troposphere, leaving the bottom boundary of the model free to simulate atmosphere-ocean exchange.

## 5. Photochemical Results

The photochemical profiles of major constituents, sulfur species, nitrogen species and hydrocarbon species in the base case model are shown in Figure 3.

The base case was modeled with the outgassing rates shown in Table 2. With these relatively low rates the base case was dominated by $CO_2$ (95%) and $N_2$ (5%). $H_2$ and CO were both in the 0.1% range, while $CH_4$ and $H_2S$ were trace gases at 0.3 ppmv and 0.1 ppbv, respectively.

Next, we increased $H_2$, sulfur ($SO_2/H_2S$), and carbon ($CO/CH_4$) outgassing to test whether or not the reducing species would effectively convert to $H_2$. Figure 4 shows the effect of increasing $H_2$ outgassing on the $H_2$ mixing ratio. As one would expect based on Eqs. [2] and [4], this produced a linear relationship. But figure 4 also shows the possible increase in atmospheric $H_2$ from the $H_2$ sources discussed in Section 3, namely, ferrous iron oxidation and serpentinization. In order to get up to 5% $H_2$ by this mechanism, the net $H_2$ sources would need to be ~80 times larger than the modern terrestrial $H_2$ outgassing rate. We can get about half of this hydrogen from direct volcanic outgassing of $H_2$ if the mantle $fO_2$ was near IW-1. Serpentinization is also a significant $H_2$ source (see Table 3). BIF deposition could contribute smaller amounts of $H_2$. If serpentinization was not as efficient as assumed here, then either volcanic outgassing rates must have been higher than we have assumed, or hydrogen escape must have been slower in order to reach the required 5% atmospheric $H_2$.

We then returned to a negligible $H_2$ outgassing rate and varied only the sulfur outgassing, keeping $SO_2$ and $H_2S$ outgassing rate in a 1:1 ratio. If the $H_2S$ is directly converted to $H_2$ we would also expect a linear relationship between the outgassing rate and $f(H_2)$.

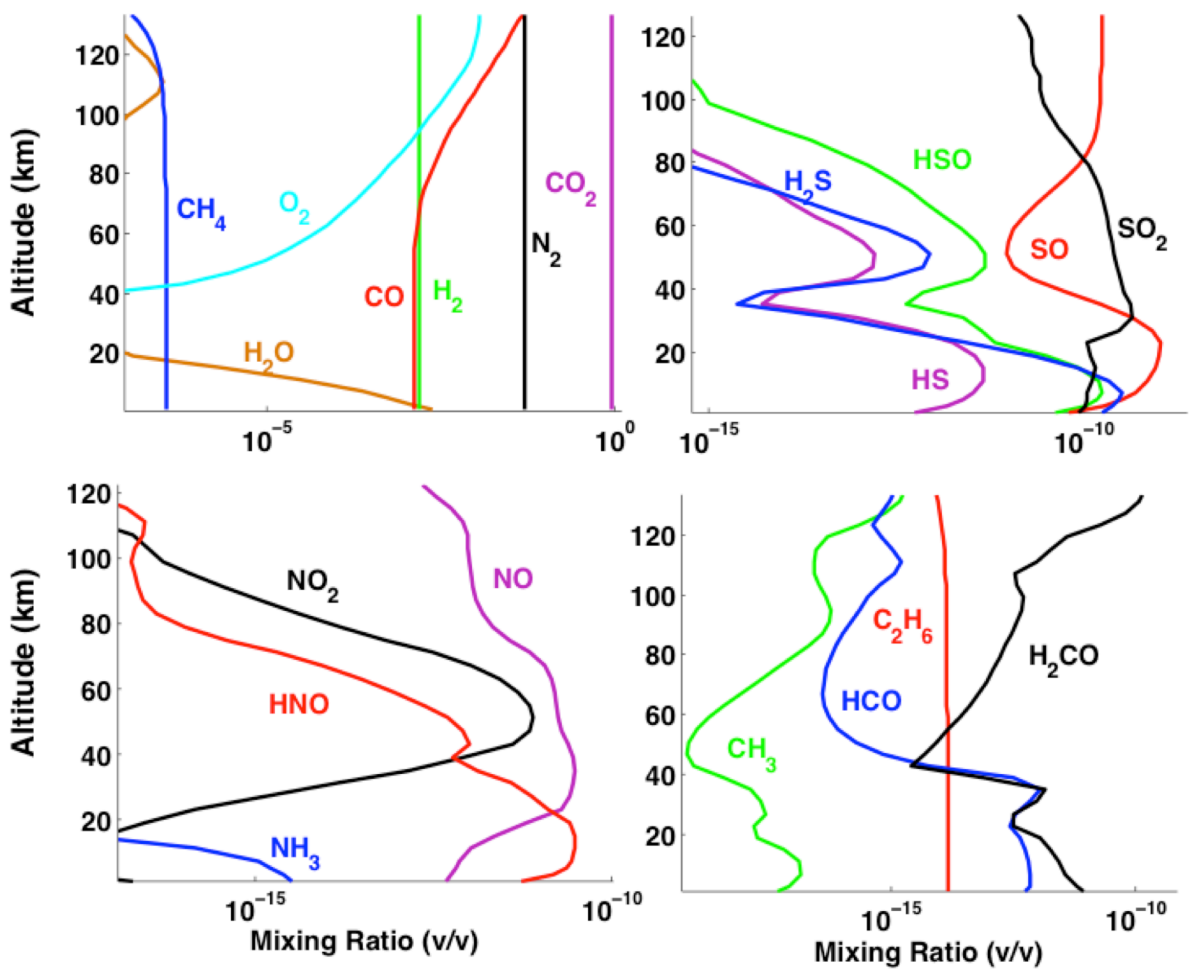

**Figure 3.** Mixing ratios of different species in the base case martian atmosphere assuming the (minimal) outgassing rates from Table 2. Major constituents are shown in the top left, sulfur species in the top right, nitrogen species in the bottom left, and less abundant hydrocarbons in the lower right.

*Table 2. Total outgassing values*

| Outgassing Rate (cm$^2$s$^{-1}$) | H$_2$ | SO$_2$ | H$_2$S | CH$_4$ | CO | NH$_3$ |
|---|---|---|---|---|---|---|
| **Base Case** | $1 \times 10^{10}$ | $5.4 \times 10^9$ | $5.4 \times 10^9$ | $5 \times 10^6$ | 0.0 | $1.7 \times 10^5$ |
| **5% H$_2$ Case** | $8 \times 10^{11}$ | $5.4 \times 10^9$ | $5.4 \times 10^9$ | $1.9 \times 10^{10}$ | $8 \times 10^9$ | $1.7 \times 10^5$ |

Instead, we found that this relationship depended on what chemistry was assumed to be occurring in the ocean. If H$_2$S was converted to H$_2$ and SO$_2$ in solution, then the stoichiometry is given by Eq. [7], and our models produce the linear relationship shown by the blue solid curve in Figure 5. By analogy with early Earth, however, it seems more likely that H$_2$S would have reacted with dissolved ferrous iron to form pyrite, FeS$_2$. The redox reaction in this case can be written as

$$2 \; H_2S + FeO \;\rightarrow\; FeS_2 + H_2O + H_2. \quad [12]$$

Based on this stoichiometry, 0.5 mole of H$_2$ should be generated for each mole of H$_2$S outgassed. This yields the blue dashed curve in Figure 5, which has a more gradual rise in the H$_2$ mixing ratio, as compared to the solid curve (~1/6th the solid line). We conclude that H$_2$S outgassing was at best a minor source of atmospheric H$_2$.

Additionally, we find that at terrestrial SO$_2$ outgassing rates ($5.4 \times 10^9$ cm$^{-2}$s$^{-1}$), H$_2$S becomes the dominant sulfur species: its concentration was ~0.1 ppbv (solid curve in figure 5), while SO$_2$ was a factor of 2-3 lower. Halevy and Head (2014) argue that Mars could have rapidly outgassed SO$_2$ over brief intervals at rates that were a few thousand times higher than modern Earth (~$10^{12}$ cm$^{-2}$s$^{-1}$), leading to 10 ppmv SO$_2$. If SO$_2$ outgassing rates were at those levels, then so were the rates of H$_2$S, and the early martian atmosphere would have been even more highly

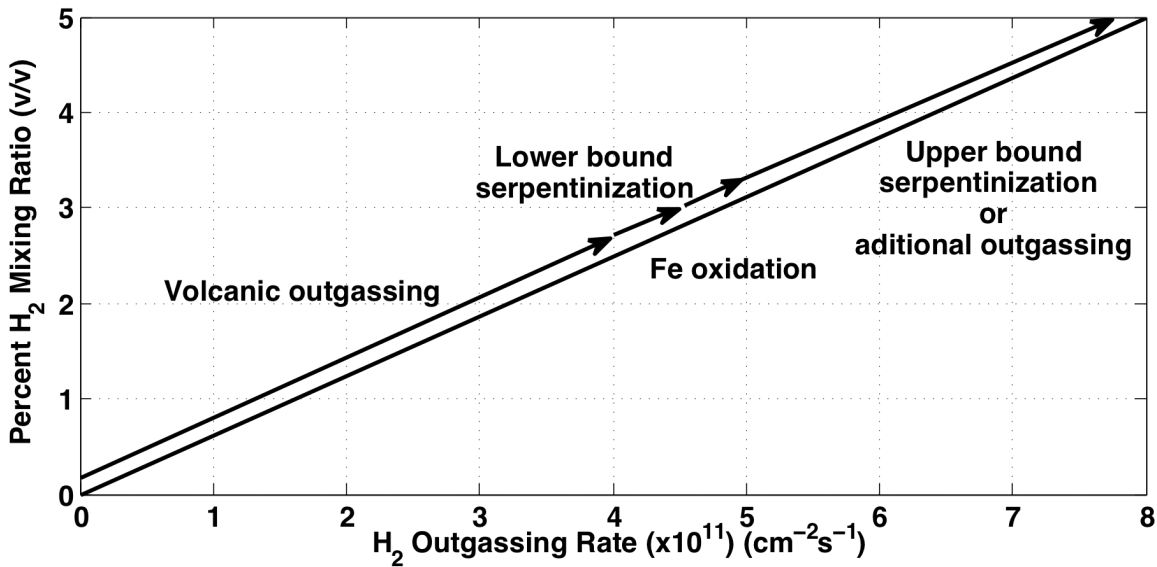

**Figure 4.** Calculation showing the effect of H$_2$ outgassing rate on H$_2$ mixing ratio, along with the different sources of H$_2$ thought to contribute to the overall outgassing rate. The escape rate is assumed to be diffusion-limited, as discussed in the text.

reducing.

We repeated this process for carbon outgassing, with the results shown in figure 6. On early Mars, carbon would have been outgassed as a combination of $CH_4$ and CO, in the ratios discussed in Section 3.1.3. As with sulfur, it is theoretically possible that some of this outgassed carbon could have been deposited in reduced form in sediments. (This is represented by the term $\Phi_{burial}(CH_2O)$ in eq. [3].) But formation of organic carbon on Earth is almost entirely biological; thus, for an abiotic early Mars, this term would probably have been small. It seems more likely that both $CH_4$ and CO were converted to $CO_2$ and $H_2$, following the stoichiometry of eqs. [8] and [9]. Therefore, carbon outgassing could have made a significant contribution to the $H_2$ concentration in the early Martian atmosphere, raising it to as high as 0.4%. But it would not have pushed the $H_2$ mixing ratio above 5% unless total carbon outgassing rates on early Mars were substantially higher than those on modern Earth.

When the carbon outgassing rate was high, the CO volume mixing ratio reached 9%, and the model atmosphere entered a regime referred to as "CO runaway" (Zahnle 1986; Kasting et al. 1983). The primary sink for CO in this situation is the flux of CO into the ocean. Because little is understood about the rate at which CO will decompose in solution, it is difficult to accurately constrain the CO deposition velocity. As a result, different authors employ different values. Kharecha et al. (2005) derived an abiotic CO deposition velocity of $10^{-9}$-$10^{-8}$ cm s$^{-1}$, based the assumption that dissolved CO equilibrates with formate, but that a small percentage of the formate is photochemically converted to acetate and is lost from the atmosphere-ocean system. If CO-consuming bacteria were present in the ocean, holding the dissolved CO concentration near zero, the CO deposition velocity would have been much higher, ~$1.2 \times 10^{-4}$ cm s$^{-1}$. Tian et al. (2014) simply logarithmically averaged these biotic and abiotic deposition velocities. There is no obvious physical justification for this assumption. Figure 7 shows a linear correlation between CO deposition velocity and CO volume mixing ratio. The loose constraint on this lower boundary condition makes it difficult to rule out the possibility of a CO-dominated atmosphere. In the limiting case where the deposition into the surface is not a sink for CO (CO deposition velocity = 0), the model atmosphere contains ~50% CO by volume. More work on the behavior of CO in solution is needed to constrain these possibilities.

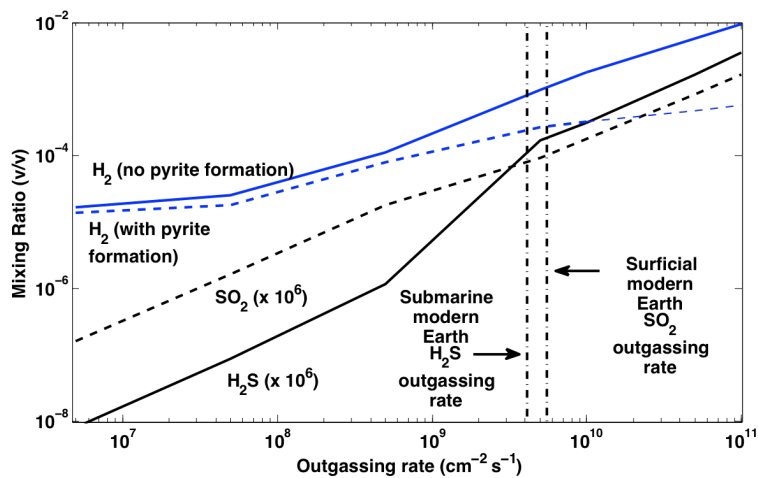

**Figure 5.** Calculation showing the effect of sulfur ($SO_2/H_2S$) outgassing rate on the atmospheric $H_2$ mixing ratio. The blue dashed curve is the $H_2$ mixing ratio under the assumption that $H_2S$ reacts to form pyrite, as seems likely. The blue solid curve shows what happens if all of the $H_2S$ dissolved in the ocean returns back to the atmosphere as a flux of $H_2$ (less likely).

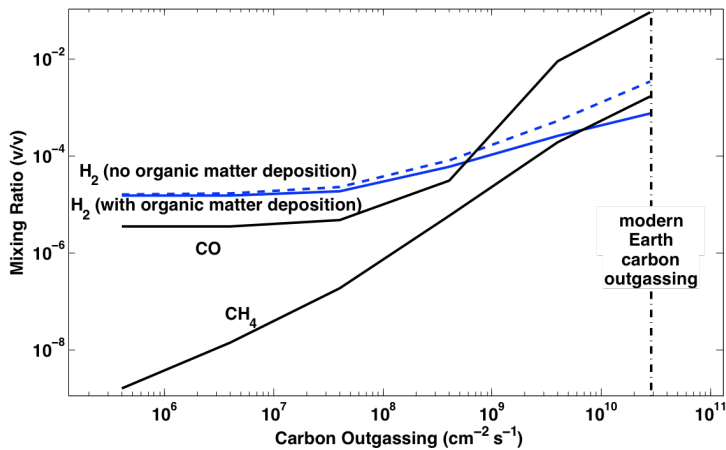

**Figure 6.** Calculation showing the effect of carbon outgassing rate on the atmospheric $H_2$ mixing ratio. At total carbon outgassing rates $> 3 \times 10^{10}$ cm$^{-2}$s$^{-1}$, the atmosphere goes into CO runaway (see discussion in text).

*Table 3. $H_2$ Sources and Respective Yields*

| $H_2$ Source | $H_2$ yield lower limit [$\times 10^{10}$ cm$^{-2}$s$^{-1}$] | $H_2$ yield upper limit [$\times 10^{10}$ cm$^{-2}$s$^{-1}$] | Value in 5% $H_2$ model [$\times 10^{10}$ cm$^{-2}$s$^{-1}$] |
|---|---|---|---|
| $H_2$ | 20 | 40 | 40 |
| S ($SO_2$ + $H_2S$) | 0.25 | 3.0 | 0.25 |
| $CH_4$ | 0.0012 | 8 | 8 |
| Serpentinization | 0.15 | 40 | 20 |
| Fe-oxide burial | 0.7 | 9 | 9 |
| Total | 21 | 100 | 77 |

Finally, figure 8 shows profiles of major atmospheric species after adding volcanic outgassing in the amounts shown in the Table 2. Table 3 shows a breakdown of the outgassing sources contributing to 5% $H_2$ compared to their upper and lower limits. In order to determine what would be required to maintain the $H_2$ greenhouse proposed by Ramirez et al. (2014), we simply assumed that the various outgassed fluxes add up to the required value. For this atmosphere, CO constituted ~9% (by volume) of this atmosphere, and the $CH_4$ mixing ratio was just under 2000 ppmv. Surprisingly, this large amount of $CH_4$ has little effect on the climate (see discussion below). Based on the discussion above, all but a small fraction of the $H_2$ in this atmosphere must have come from direct $H_2$ sources, such as $H_2$ outgassing, ferrous iron oxidation, and serpentinization.

## 6. Discussion

### 6.1 Greenhouse warming by gases other than $CO_2$ and $H_2$

#### 6.1.1 CO & $CH_4$

A 3-bar, $CO_2$-dominated atmosphere with 5% $H_2$ could have warmed the early Martian surface. But our high-outgassing atmosphere also contained almost 2000 ppmv of $CH_4$, which is considered to be a strong greenhouse gas, along with ~10% CO. CO has an absorption band in the 5μm region, far into the Wien tail of a blackbody with an effective temperature 235 K. Therefore, despite its high concentration, the effect of CO on climate is limited to pressure-broadening of gaseous absorption by other species and Rayleigh scattering of incident solar radiation. (We tested this just to make sure by deriving *k*-coefficients for CO and including it in our climate model. The effect was negligible.)

Methane's effect on climate in a dense, $CO_2$-dominated martian paleoatmosphere has previously been explored (Ramirez et al. 2014; Byrne & Goldblatt 2014). Both groups find little to no warming from $CH_4$. Methane has a strong absorption band at 7.7 μm that is important in warming Earth's climate. However, near-infrared absorption of incoming solar radiation by $CH_4$ in the upper atmosphere produces stratospheric inversions that counteract this greenhouse warming. This stratospheric warming is particularly pronounced when new $CH_4$ absorption coefficients derived from the HITRAN2012 database are used (Byrne & Goldblatt 2014). Collision-induced absorption (CIA) from $N_2$-$CH_4$ interactions is significant in the 200-400 $cm^{-1}$ region. However, because the strong water vapor pure rotation band also absorbs in that same region (200-400 $cm^{-1}$), the warm atmosphere would be opaque at those wavelengths and CIA does little to help sustain this (Buser et al. 2004; Ramirez et. al. 2014). As a result, the additional warming produced by $CH_4$ in this atmospheres modeled here should be negligible. An important caveat is our explicit assumption that $CO_2$ broadening of $CH_4$ is no more efficient than broadening by $N_2$. Although experimental data for $CO_2$-$CH_4$ CIA interactions are currently unavailable, we should not rule out methane as a potential source of warming if it is later found that $CO_2$ is a better foreign broadening agent than $N_2$.

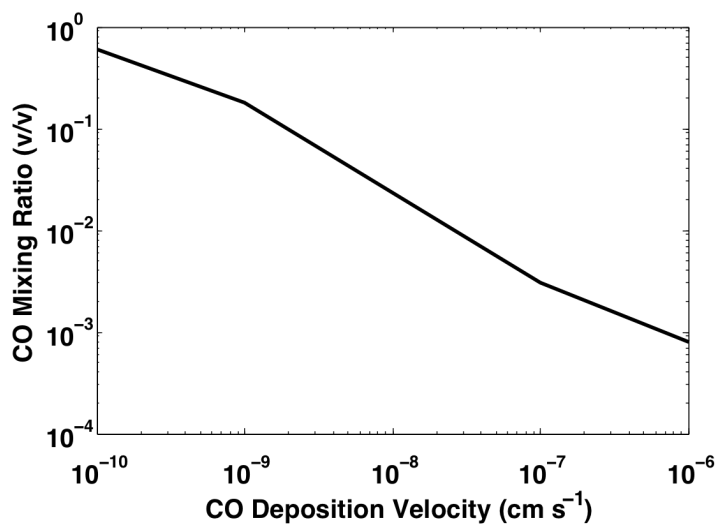

**Figure 7.** CO volume mixing ratio as a function of assumed deposition velocity. The atmospheric profile in Figure 8 was used for all calculations. The loose constraint on deposition velocity prevents us from determining precise values for CO

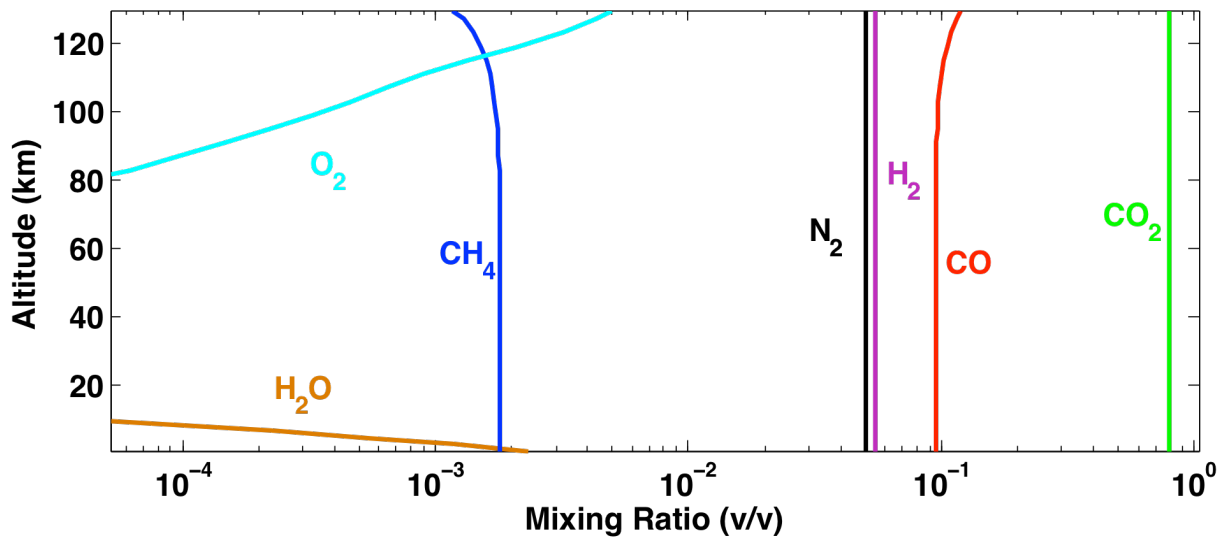

**Figure 8.** 1-D photochemical model results showing volume mixing ratio as a function of altitude. This is the base case early Martian atmosphere (Figure 2) after adding volcanic outgassing and balancing the redox budget for the combined ocean-atmosphere system. Here we are optimistic, and add in enough $H_2$ outgassing to attain a 5% $H_2$ atmosphere.

*6.1.2 SO$_2$*

As mentioned earlier, Halevy and Head (2014) suggested that warm periods lasting 10-100 years could have been produced by short, episodic bursts of $SO_2$. Specifically, they assumed long-term globally averaged $SO_2$ outgassing rates on the order of $10^{10}$ cm$^{-2}$s$^{-1}$, with outgassing events on the order of $10^{12}$ cm$^{-2}$s$^{-1}$ every 1,000-10,000 years. This leads to $SO_2$ concentrations ranging from 0.5-2 ppmv in their model. By comparison, our assumed $SO_2$ outgassing rate for both the base-case and high-$H_2$ models is $5.4 \times 10^9$ cm$^{-2}$s$^{-1}$ (Table 2), or just over half that of Halevy and Head (2014), but our calculated $SO_2$ concentrations are ~3 orders of magnitude smaller. The difference is caused by our assumption that early Mars was wet and warm and that an ocean—or at least several large seas—was present at its surface. Both rainout and surface deposition are therefore important loss processes in our model, whereas the (much longer) lifetime of $SO_2$ in the Halevy and Head model is set by its rate of photochemical oxidation to $H_2SO_4$. Their model would predict smaller, shorter-lived temperature increases if rainout and surface deposition of $SO_2$ were included.

We note that sporadic, high-volume input of $SO_2$, as suggested by Halevy and Head, should have been accompanied by high-volume input of $H_2$. This should not have had a great impact on the atmospheric $H_2$ concentration, however, because the lifetime of $H_2$ in one of our high-$H_2$ atmospheres is close to half a million years. (This can be readily calculated by dividing the column mixing ratio of $H_2$ by the diffusion-limited escape rate given by eq. [2].) In the Halevy and Head model, slow outgassing during the long periods of relative quiescence dominates the total volatile input, and the same should be true of $H_2$. 100 years of $H_2$ outgassing at 100 times the normal rate would have increased atmospheric $H_2$ concentrations by only a few percent. So, a spiky volcanic outgassing history for early Mars would not alter our hypothesis to any great extent.

### 6.2 S-MIF Signal Implications

A concern with the proposed $H_2$-dominated atmosphere is that it would have eliminated the oxidized sulfur exit channels, and thereby have precluded any sulfur mass independent fractionation (MIF) from being recorded in the rock record. This signal can be measured as $\Delta^{33}S$, the deviation of the $^{33}S/^{32}S$ ratio from the fractionation line defined by $^{34}S$ and $^{32}S$. A recent analysis of 40 Martian meteorites reveals sulfur isotopes indicative of mass-independent fractionation (MIF) in a variety of protolithic ages - ALH 84001, the nakhlites, Chassigny and six shergottites (Franz et al. 2014). The only way to preserve such a S-MIF signal is if sulfur is distributed amongst two or more different species as they rain out of the atmosphere (Pavlov & Kasting 2002). Our simulations predict sulfur would have exited the atmosphere in at least three different exit channels, HSO, $SO_2$, $H_2S$ (see Figure 9) even with 5% $H_2$. This suggests that an atmosphere containing 5% $H_2$ could still produce and record a measureable S-MIF signal.

### 6.3 D/H ratios, hydrogen escape rates, and initial water inventories

The recent paper by Villanueva et al. (2015) provides additional support for the idea that early Mars was warm and wet. These authors looked at deuterium/hydrogen (D/H) ratios in various water vapor masses across the martian surface and estimated an average enrichment of ~8 for their source regions (martian ice) relative to terrestrial seawater. When combined with an estimated modern $H_2O$ inventory of 21 m GEL (global equivalent layer) in the polar layered deposits, an estimated initial D/H enrichment of 1.275 relative to seawater, and an assumed fractionation factor, $f = 0.02$, for escape of D relative to H, this yields a global equivalent water layer of 137 m for early Mars. The relevant mathematical relation is

$$\frac{M_p}{M_c} = \left(\frac{I_p}{I_c}\right)^{1/(1-f)}.$$

[13]

Here, $M_p$ and $M_c$ are the ancient and current water reservoir sizes, respectively and $I_p$ and $I_c$ are the ancient and current D/H ratios.

137 m of water may sound like a lot, but in reality it is just a lower bound because the assumed fractionation factor, from Krasnopolsky et al. 1998, is only appropriate for the modern (highly tenuous) martian upper atmosphere in which nonthermal hydrogen escape processes predominate. If the early martian atmosphere was rich in $H_2$, as postulated here, hydrogen escape would have been hydrodynamic, and D would have been dragged off along with H, thereby increasing the fractionation factor, $f$. We can estimate $f$ if we assume that $H_2$ was escaping at the rate of $8 \times 10^{11}$ $cm^{-2}s^{-1}$ required to maintain a 5% $H_2$ atmosphere. We assume here that hydrodynamic escape was efficient enough to keep up with the diffusion limit. The fractionation factor for hydrodynamic escape is given by (Hunten et al. 1987, eq. 17)

$$f \equiv \frac{F_2/X_2}{F_1/X_1} = \frac{m_c - m_2}{m_c - m_1} \qquad (14)$$

Our notation is slightly different from Hunten et al., and our fractionation factor, $f$, is related to their factor, $y$, by $f = \dfrac{1}{1+y}$. (The '$y$' notation is convenient for isotopes of heavy elements that differ in mass by only a small percentage, whereas the $f$ notation is preferred for isotopes of light elements like hydrogen that have large mass differences.) Here, $F_1$, $X_1$, and $m_1$ are the escape rate, mixing ratio, and molecular mass of the lighter species ($H_2$), $F_2$, $X_2$, and $m_2$ are the equivalent quantities for the heavier species (HD), and $m_c$ is the crossover mass, given by (Hunten et al., 1987, eq. 16)

$$m_c = m_1 + \frac{kTF_1}{bgX_1} \qquad [15]$$

Here, $k$ is Boltzmann's constant, $T$ is temperature, $X_1$ is the mole fraction of $H_2$ ($\cong 1$), $g$ (= 373 cm $s^2$) is gravity, and $b$ (=$1.76 \times 10^{19}$ $cm^{-1}s^{-1}$) is the diffusion constant between $H_2$ and HD (Banks and Kockarts, 1973, v. 2, eq. (15.29)). We can simplify this expression by dividing through by the mass of a hydrogen atom, $m_H$, and letting $H_H = \dfrac{kT}{m_H g}$ be the scale height of atomic hydrogen. Then, in atomic mass units, eq. (15) becomes

$$M_c = M_1 + \frac{F_1}{b/H_H} \qquad (16)$$

If we take $T = 160$ K, then $H_H \cong 3.5 \times 10^7$ cm and $b/H_H \cong 5 \times 10^{11}$ cm$^{-2}$s$^{-1}$. For $F_1 = 8 \times 10^{11}$ cm$^{-2}$s$^{-1}$, we get $M_c \cong 3.6$ amu, and from eq. (14), $f \cong 0.4$. Then, if we take the other parameters to be the same as those assumed by Villanueva et al. (2015), Eq. (13) yields an initial water inventory of ~450 m, or over three times their published estimate. Higher hydrogen escape fluxes would increase this value even further, following the nonlinear relationships expressed by eqs. (13-16). We conclude that high measured D/H ratios on present Mars are consistent with a relatively deep global ocean and an $H_2$-rich atmosphere on early Mars. They do not require it, however, as these same high D/H ratios can be produced by loss of lesser amounts of water by mechanisms involving lower fractionation factors.

### 6.4 Tests for higher $H_2$ outgassing rates

As discussed above, a 5% $H_2$ atmosphere is possible if $H_2$ outgassing rates on ancient Mars were $8 \times 10^{11}$ $H_2$ molecules cm$^{-2}$s$^{-1}$; however, the assumed outgassing rates in our standard simulations only provide ~$5 \times 10^{11}$ $H_2$ molecules cm$^{-2}$s$^{-1}$, if serpentinization did not generate significant $H_2$. If Mars' early atmosphere was hydrogen-rich, at least one of our estimated $H_2$ outgassing sources must be too low, or else $H_2$ must have escaped at less than the diffusion-limited rate. The escape rate can, in principle, be investigated by constructing sophisticated theoretical models of hydrodynamic escape. This remains as work to be done. Some empirical tests of hydrogen outgassing and escape rates may be possible, either using rover measurements or by analyses of samples returned from the Martian surface. For exampled, Curiosity has already made a significant impact on D/H ratio measurements. D/H was measured in a ~3-billion-year-old (Farley et al., 2014) mudstone at 3.0 times the ratio in standard mean ocean water (Mahaffy et al. 2015). This value is half the D/H ratio of present Mars' atmosphere, which is consistent with continued escape of hydrogen throughout Mars' history. By making similar measurements as Curiosity moves up-section in Gale Crater, and comparing these measurements to existing measurements of Martian meteorites, a history of martian D/H ratios can be constructed, and from this some information regarding the H escape rate can be inferred. Of course, getting detailed information is complicated because the fractionation between D and H during escape depends on both the rate and mechanism of the escape process. More robust tests are outlined briefly below.

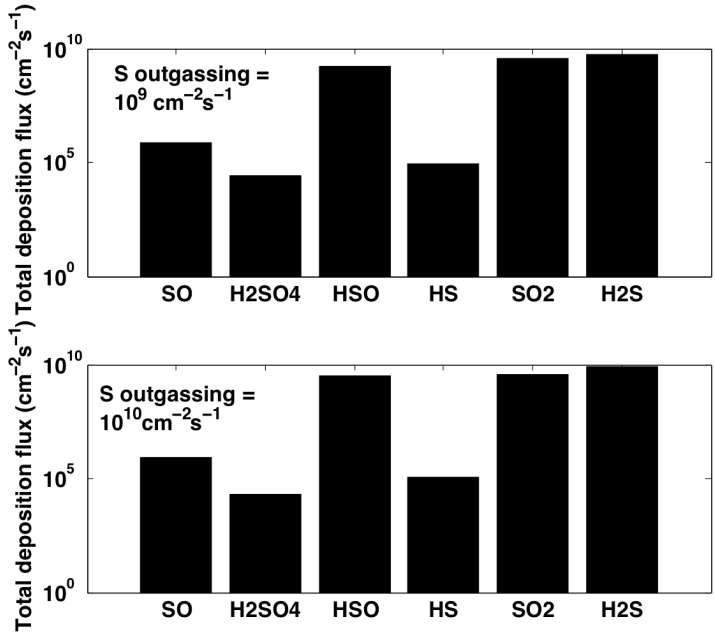

**Figure 9.** Total removal rate (rainout + surface deposition) for low (top) and high (bottom) cases of sulfur outgassing in an early Martian atmosphere. This shows three quantitatively important pathways that should allow for the preservation of a sulfur isotope MIF signal.

### 6.4.1 Analyses of ancient Martian mantle redox state

The single greatest source of uncertainty in our $H_2$ outgassing budgets is our knowledge of the redox state of the ancient martian mantle. The majority of martian meteorites have a mantle oxygen fugacity near IW+1. But if even part of the story we have outlined here is true, then the redox state of the martian mantle must have evolved with time as $H_2O$ was subducted and hydrogen was outgassed as $H_2$, leaving oxygen behind. Other authors, too, have postulated that the redox state of the martian mantle evolved over time (Righter et al. 2008). An initial redox state of IW-1, or lower, is consistent with a more reducing composition during core formation and with the low measured $fO_2$ of ALH 84001 (Warren & Gregory 1996; Steele et al., 2012). The $fO_2$ for ALH 84001 is closer to IW-1, a value that would provide a half the $H_2$ flux needed to maintain a 5% $H_2$ atmosphere, given Earth-like outgassing rates. So, even at this point, our model requires twice Earth outgassing rates (or slower $H_2$ escape). As the mantle $fO_2$ increased, even higher outgassing rates would have been needed to maintain 5% H2. Thus, the end of the warm wet period could have been brought about either by declining outgassing rates or by progressive mantle oxidation.

Such an evolution of mantle redox state would be broadly consistent with other assumptions in our conceptual model. Deposition of oxidized iron in BIFs and subsequent subduction of these sediments would have deposited additional O in the mantle. We should note that a similar process of progressive mantle oxidation has been suggested for early Earth (Kasting 1993) but has since been largely ruled out. On Earth, these processes evidently did *not* result in a secular change in mantle redox state, presumably because Earth's mantle was already oxidized up to near the QFM buffer during or shortly after accretion (Wade & Wood 2005; Frost & McCammon 2008). The proposed oxidation process involves disproportionation of ferrous iron at high pressures in Earth's lower mantle—a process that might not have occurred on a smaller planet. If the martian mantle started out with an $fO_2$ near IW, then small additions of oxygen could conceivably have oxidized it more significantly over time. Indeed, the observed spread in $fO_2$ values of martian meteorites up to values approaching QFM suggests that mantle oxidation did occur (Stanley et al. 2011). We take this as indirect support for our hypothesis.

That said, the existing data do not allow us to draw firm conclusions about the early redox evolution of the martian mantle. The only measurement of redox state during this early phase of the planet's history comes from analyses of ALH 84001. Given the large spread in measured $fO_2$ values of younger materials, this leaves great uncertainties in the redox state at that time. Ideally, one would like to have a suite of redox measurements at multiple points in martian history, to account for the spread in the data and to constrain the temporal evolution. This is not feasible in the near future, but could happen over the coming decades with an extensive sample return campaign.

In the meantime, contextual information on this period of martian history could be obtained from Curiosity, ExoMars, and the Mars 2020 lander. Such contextual information on the early surface evolution of Mars is one of the main goals of the Curiosity mission. Curiosity has already dated sedimentary rocks on the martian surface (Farley et al. 2014), placing an age on the deposition of the lacustrine sediments in Gale Crater (Grotzinger et al. 2013). Making subsequent time-stamped measurements would help place the rest of the results from Curiosity in an absolute historical context that could be compared to future mantle redox measurements.

This context would be augmented by qualitative estimates of the redox state of mantle-derived materials. These can be made through this measurement of the relative abundances of redox-sensitive trace elements such as Fe, for example with MSL's ChemCam. Similar instrumentation was present on the MER rovers, and planned for future rovers, so it may be possible to stitch together this history, albeit without the quantitative dating capabilities of Curiosity.

*6.4.2 Analyses of Fe-oxide rich sedimentary rocks*
Contextual information on the co-evolution of the Martian atmosphere and mantle may also come through analyses of ancient Fe-oxide rich layers. Although we do not expect a huge $H_2$ outgassing contribution from Fe-oxide deposition (see sections 3.2-3.3), the presence of Fe-oxide rich sedimentary layers is consistent with an $H_2$-rich atmosphere. Such layers have an analog in the banded iron-formations (BIFs) on ancient Earth. These are thought to have required an anoxic deep ocean so that ferrous iron could be transported over long distances to upwelling regions where the BIFs formed (Holland 1973). Therefore, charting and dating the presence/absence of such layers could provide critical information on the long-term evolution of the redox state of the surface, just as the presence/absence of BIF's has had a huge impact on our assessment of the redox history of Earth's surface (Holland & Trendall 1985).

By measuring the deposition rates of the Fe-oxides, it would also be possible to estimate the $H_2$ outgassing rate they could have provided, both locally and, by extrapolation, globally. The Fe concentration of the samples can be measured, and the deposition times can in theory be calculated by dating the sedimentary layers on Mars (Farley et al., 2014).

However, the uncertainties of the dating measurements (+/- 0.35 Ga) are significantly longer than the timescales of deposition, and many of the concentration measurements will be qualitative in nature. Further, extrapolation will be difficult unless the size of the lake above the floor of Gale Crater can be estimated and compared to estimates of the contemporaneous global reservoir. That said, even qualitative information on the Fe-oxide deposition rate and its evolution over time would be useful. This could lead to order-of-magnitude inferences on the Fe flux from the subsurface, and the rate at which the mantle was being oxidized through Fe-oxide burial.

## 7. Conclusions

The idea that early Mars was warm and wet for prolonged time intervals, millions of years or more, is consistent with new data from the MSL mission. The only published mechanism that appears capable of maintaining such conditions for extended periods is the greenhouse effect of a $CO_2$-$H_2$ atmosphere. About 3 bar of $CO_2$ and 5 percent or more $H_2$ is required to produce global mean surface temperatures above freezing. Maintaining $H_2$ at this level is challenging but does not appear to be out of the question. Direct volcanic outgassing of $H_2$ from a highly reduced early martian mantle was probably the largest source of $H_2$. Recycling of volatiles between the surface and the interior, as happens on Earth because of plate tectonics, would likely have been needed to provide this $H_2$, as rates of juvenile outgassing are small. Additional $H_2$ could have been provided by photochemical oxidation of outgassed $CH_4$ and $H_2S$ and by processes such as serpentinization and deposition of banded iron-formations. However, none of these sources are sufficient unless: i) the ancient Martian mantle was significantly more reduced than today, ii) volcanic outgassing rates were substantially higher than those on modern Earth or iii) hydrogen escaped to space more slowly than the diffusion limit. Some combination of these three mechanisms could also work.

Recycling of water through the mantle followed by outgassing of $H_2$ should have oxidized the mantle over time. Such oxidation is consistent with the observed spread in $fO_2$ values of SNC meteorites from as low as IW-1 up to near QFM. Additional tests of the $H_2$ greenhouse hypothesis may be provided by MSL and by future missions. MSL itself could look for evidence of banded iron-formations and changes in D/H ratios that might indicate hydrogen loss, as well as providing qualitative and contextual information on the redox evolution of the martian mantle. Future sample return missions could look for quantitative evidence of secular mantle oxidation over time. Improved numerical models of hydrodynamic escape could shed light on the escape rate of H over time. Additional 3-D climate modeling work would also be useful to better constrain the amounts of rainfall and surface runoff that could have been maintained by this model and by its competitors, and compare these results to the lacustrine and fluvial deposits in Gale Crater and across the planet.

## Acknowledgements

This material is based upon work supported by the National Science Foundation under Grant No. DGE1255832 to N. Batalha. Any opinions, findings, and conclusions or recommendations expressed in this material are those of the author(s) and do not necessarily reflect the views of the National Science Foundation. JFK acknowledges support from NASA's Exobiology and Astrobiology programs. R.R. acknowledges support from the Simon's Foundation (SCOL 290357, L.K.) and the Carl Sagan Institute.